\documentclass[12pt]{iopart}
\usepackage{epsf}

\begin{document}

\title{A deformable elastic membrane embedded in a lattice Boltzmann fluid}

\author{S V Lishchuk and C M Care}

\date{}

\address{
Materials and Engineering Research Institute, Sheffield Hallam University,\\
Howard Street, Sheffield S1~1WB, United Kingdom
}

\begin{abstract}
A method is described for embedding a deformable, elastic, membrane
within a lattice Boltzmann fluid. The membrane is represented by a
set of massless points which advect with the fluid and which impose
forces on the fluid which are derived from a free energy functional
with a value which is dependent upon the geometric properties of the
membrane. The method is validated in two dimensions with a free
energy functional which imposes the constraint of constant membrane
length, constant enclosed area, a bending rigidity and a preferred
curvature. The method is shown to recover the expected equilibrium
shape in the absence of flow and deformation in the presence of an
applied shear flow. The method may have applications in a number of
mesoscopic simulations, including discrete models of blood cells.
\end{abstract}



\maketitle


\section{\label{sec:Introduction}Introduction}

A number of biological objects can be represented as vesicles formed
by polymerised membranes (Alberts \etal 2002, Lipowsky 1991).
While the conformations of such vesicles are determined by the
elastic properties of the membrane (Lipowsky 1991), their
dynamics in a flowing fluid is altered by the flow of the fluid
outside and inside vesicles. Both factors should be taken into
account in simulation of the membranes immersed into fluid host.

The lattice Boltzmann (LB) method (Succi 2001) provides a
convenient method for introducing fluid flow in the presence of
boundaries. An extension of the LB method to model fluid membranes
in which the molecules rapidly diffuse within the membrane has been
reported (Stelitano and Rothman 2000). However, this method cannot be
straightforwardly generalised to the polymerised membranes since it
does not allow for the extensional elastic properties which would be
associated with such a membrane. An alternative approach is to model
the membrane as a geometric object immersed into the LB fluid and
this is the method adopted in this paper. A similar approach has
been developed for the simulation of a polymer chain in an LB
solvent (Ahlrichs and D\"unweg 1998).

The aim of the present paper is to develop an LB method for the
polymerised membranes. The main purpose of the paper is
methodological and we therefore demonstrate the effectiveness of the
method for a simple two-dimensional case. However, we discuss the
generalisation to three dimensions in section~\ref{sec:Conclusion}.
It is important to note that the motivation for the work in this
paper is ultimately to develop models for flows in which there are
embedded a large number of deformable membranes; this will form the
basis of modelling blood  flow at the veinule scale.  For this
reason, the finer details of the membrane properties are ignored
({\it eg} membrane dissipation).

The paper is organised as follows. In section~\ref{sec:LB} the
method for introducing the membrane into the LB scheme is described.
The explicit expressions for the forces arising from the membrane
are derived in section~\ref{sec:Force}. The results of the
simulations of the relaxation of a closed membrane to its
equilibrium state without and with shear flow are presented in
section~\ref{sec:Results}. Conclusions and possible extensions and
applications are discussed in section~\ref{sec:Conclusion}.


\section{\label{sec:LB}Lattice Boltzmann}

The basics of the LB method have been described in the literature
(Succi 2001). The fluid in the LB method is considered as a
field of the population densities $f_i(\mathbf r,t)$, which indicate
the amount of fluid present at the lattice site $\mathbf r$ at the
discrete time $t$ and moving with the velocity $c_i$ associated with
the $i$-th lattice direction. The models with $n$ velocities on a
simple cubic lattice of dimension $d$ are usually referred to as
D$d$Q$n$. The LGBK algorithm may be represented by the equation
\begin{equation}
f_{i}(\mathbf{r}+\mathbf{c}_{i}\delta_{t},t+\delta_{t})
=f_{i}(\mathbf{r},t)+\frac{1}{\tau}(f_{i}^{(0)}-f_{i})+G_{i},
\label{eq:LB}
\end{equation}
where $\delta_{t}$ represents the time step, $\tau$ controls the
kinematic shear viscosity of the lattice fluid through the relation
\begin{equation}
\nu=\frac{2\tau-1}{6}\delta_{t},
\label{eq:viscosity}
\end{equation}
and the \lq forcing' term
\begin{equation}
G_{i}=\frac{1}{c_{s}^{2}}t_ic_{i\mu}F_{\mu},
\label{eq:LBforce}
\end{equation}
may be used to impress an external force $F_{\mu}$, where $c_s$ is
the speed of sound and the $t_i$ are determined to achieve isotropy
of the fourth-order tensor of velocities and Galilean invariance.
Note that the expressions (\ref{eq:LB})--(\ref{eq:LBforce}) are
independent of the spatial dimension.

Velocity moments give the lattice fluid's density and momenta through
\begin{equation}
\rho=\sum_{i}f_{i}=\sum_{i}f_{i}^{(0)},
\label{eq:LBdensity}
\end{equation}
\begin{equation}
\rho\mathbf{v}=\sum_{i}\mathbf{c}_{i}f_{i}=\sum_{i}\mathbf{c}_{i}f_{i}^{(0)},
\label{eq:LBvelocity}
\end{equation}
where the equilibrium distribution function $f_{i}^{(0)}$ is
\begin{equation}
f_{i}^{(0)}(\rho,\mathbf{v})=
t_i\rho\left[1+\frac{\mathbf{v}\cdot\mathbf{c}_{i}}{c_{s}^{2}}+
\frac{|\mathbf{v}|^{2}}{2c_{s}^{2}}+\frac{(\mathbf{v}\cdot\mathbf{c}_{i})^{2}}{2c_{s}^{4}}\right].
\label{eq:LBf0}
\end{equation}
The form of the equilibrium distribution function (\ref{eq:LBf0})
ensures that relations (\ref{eq:LBdensity},\ref{eq:LBvelocity})
are satisfied and also determines the nonviscous part of the momentum-flux
tensor of the lattice fluid,
\begin{equation}
\Pi_{\alpha\beta}^{(0)}=\sum_{i}f_{i}^{(0)}c_{i\alpha}c_{i\beta}=
c_{s}^{2}\rho\delta_{\alpha\beta}+\rho v_{\alpha}v_{\beta}.
\end{equation}
The membrane is represented by a discrete set of points
corresponding to the equidistant values of a parameter $s$. The
difference $\Delta s=s_{i+1}-s_{i}$ between the values of the
parameter $s$ for the consecutive points is chosen to be comparable
with the distance between LB nodes.

The membrane is treated purely as a geometric object and, contrary
to (Ahlrichs and D\"unweg 1998), the points of the membrane always move
with the velocity of the underlying LB fluid determined from
equation (\ref{eq:LBvelocity}). Generally, the position of the
points does not coincide with the location of LB nodes, so a
weighted average is used based on the velocities at the nodes which
bound the lattice primitive cell in which the point is situated. The
weight of the contribution from each node is taken to be
proportional to the distance of the node from the membrane point
under consideration. The geometric properties of the membrane
determine the force that is in turn applied to the LB fluid
according to equation (\ref{eq:LBforce}). Again, as the location of
the point generally does not coincide with nodes, the force is
distributed amongst the same set of nodes in accordance with their
distance from the point at which the force originates. The value of
the force is determined in section~\ref{sec:Force}.

It should be noted that a membrane represented in this way will not
strictly conserve its internal mass, and will therefore be slightly
permeable.  However, since we are modelling polymerised membranes
which are themselves weakly permeable, this is not considered to be
a serious defect in the approach.


\section{\label{sec:Force}Forces}

We describe the shape of the membrane by the vector function
$\mathbf{r}(s)$ having the components $x(s)$ and $y(s)$. We choose
the parameter $s$ so that for the undeformed membrane (without
internal stress) $s$ coincides with the arc length parameter $l$ of
the curve. The parameter $s$ spans the interval
$\left(0,L_{0}\right)$, $L_{0}$ being the equilibrium length of the
membrane. In general, the length of the membrane is determined by
the expression
\begin{equation}
L=\int_{L}dl=\int_{0}^{L_{0}}u(s)ds.
\end{equation}
Here $dl=u(s)ds$ is the length element,
\begin{equation}
u(s)=\sqrt{x'^{2}(s)+y'^{2}(s)}.
\end{equation}
If the membrane is stretched/compressed, the free energy increases.
The excess free energy is
\begin{equation}
A_{L}[\mathbf{r}(s)]=\frac{\alpha}{2}\int_{0}^{L_{0}}\left(u(s)-1\right)^{2}ds,
\label{eq:AL}
\end{equation}
$\alpha$ being the membrane compressibility. The free energy arising
from the bending elasticity of the membrane can be taken in form
(Canham 1970, Helfrich 1973, Evans 1974):
\begin{equation}
A_{K}=\frac{\kappa}{2}\int_{L}\left(K(\mathbf{r})-K_{0}\right)^{2}dl,
\end{equation}
$K(\mathbf{r})$ being the curvature, $K_{0}$ being the preferred
curvature and $\kappa$ being the bending rigidity coefficient. The
curvature can be represented as a function of the parameter $s$
\begin{equation}
K(s)=\frac{x'(s)y''(s)-y'(s)x''(s)}{u^{3}(s)}.
\end{equation}
and hence we can write
\begin{equation}
A_{K}[\mathbf{r}(s)]=\frac{\kappa}{2}\int_{0}^{L_{0}}\left(K(s)-K_{0}\right)^{2}u(s)ds.
\label{eq:AK}
\end{equation}
The contribution to the free energy due to the surface tension is
$A_{S}=\sigma L$, or
\begin{equation}
A_{S}=\sigma\int_{0}^{L_{0}}u(s)ds,
\label{eq:AS}
\end{equation}
$\sigma$ being the surface tension coefficient. The fluid is assumed
to be compressible and the equilibrium two-dimensional \lq volume'
of the fluid enclosed by the membrane is taken to be $V_{0}$. The
excess pressure is
\begin{equation}
p=-\beta\left(V[\mathbf{r}(s)]/V_{0}-1\right),
\end{equation}
where $\beta$ is the fluid compressibility, and the \lq volume' of
the droplet $V[\mathbf{r}(t)]$ is the functional of the membrane
shape
\begin{equation}
V[\mathbf{r}(s)]=\int_{0}^{L_{0}}y(s)x'(s)ds.
\end{equation}
The free energy due to the fluid compressibility is
\begin{equation}
A_{V}=\int_{V_0}^{V}pdV
\end{equation}
or, after integration,
\begin{equation}
A_{V}[\mathbf{r}(s)]=\frac{\beta}{2V_{0}}\left(V[\mathbf{r}(s)]-V_{0}\right)^{2}.
\label{eq:AV}
\end{equation}

As a result, the free energy of the interface can be represented as
a following functional of the membrane shape:
\begin{equation}
A[\mathbf{r}(s)]=A_{L}[\mathbf{r}(s)]+A_{K}[\mathbf{r}(s)]+A_{S}[\mathbf{r}(s)]+A_{V}[\mathbf{r}(s)],
\end{equation}
where $A_{L}[\mathbf{r}(s)]$, $A_{K}[\mathbf{r}(s)]$, $A_{K}[\mathbf{r}(s)]$
and $A_{V}[\mathbf{r}(s)]$ are given by formulas (\ref{eq:AL}),
(\ref{eq:AK}), (\ref{eq:AS}), and (\ref{eq:AV}), correspondingly.

The force $\mathbf{F}$ from the element $dl$ of the membrane is
found as the variational derivative of the free energy,
$\mathbf{F}(s)=-\delta A[\mathbf{r}(s)]/\delta\mathbf{r}(s)$,
and can be represented in form
\begin{equation}
\mathbf{F}(s)=\mathbf{F}_{L}(s)+\mathbf{F}_{K}(s)+\mathbf{F}_{S}(s)+\mathbf{F}_{V}(s)
\label{eq:F}
\end{equation}
with the components
\begin{equation}
\mathbf{F}_{L}(s)=-\alpha\left(K(s)\mathbf{n}(s)+\partial^{2}\mathbf{r}/\partial s^{2}\right),
\label{eq:FL}
\end{equation}
\begin{equation}
\mathbf{F}_{K}(s)=\kappa\left[\frac12K(s)\left(K(s)^{2}-K_{0}^{2}\right)+\frac{d^{2}K(s)}{dl^{2}}\right]\mathbf{n}(s),
\label{eq:FK}
\end{equation}
\begin{equation}
\mathbf{F}_{S}(s)=-\sigma K(s)\mathbf{n}(s),
\label{eq:FS}
\end{equation}
\begin{equation}
\mathbf{F}_{V}(s)=-\beta\left(V[\mathbf{r}(s)]/V_{0}-1\right)\mathbf{n}(s),
\label{eq:FV}
\end{equation}
corresponding to the membrane compressibility, bending elasticity,
surface tension and fluid compressibility. In formulas
(\ref{eq:FL}--\ref{eq:FV}), $\mathbf{n}(s)$ is the unit vector
normal to the membrane with the components $n_x(s)=y'(s)/u(s)$ and
$n_y(s)=-x'(s)/u(s)$. The expression for the force (\ref{eq:FK}) is
equivalent to that in (Stelitano and Rothman 2000) after it has been
noted that there is an error in the quoted result  which arises
because the authors do not correctly account for the change in the
metric tensor of the surface during the minimisation of the free
energy (Lishchuk and Care 2005). Equation (\ref{eq:FK}) also includes an
additional contribution due to the preferred curvature $K_{0}$.

The functions $x(s)$ and $y(s)$ can be approximated by polynomials.
We use the second order polynomials,
\begin{eqnarray*}
x(s) & = & a_{0}+a_{1}s+a_{2}s^{2},\\
y(s) & = & b_{0}+b_{1}s+b_{2}s^{2}.
\end{eqnarray*}
In order to determine the coefficients $a_{k}$ and $b_{k}$
$\{k=0,1,2\}$ at the $i$-th point of the discretised membrane, we
require the values of the functions at this and two neighbouring
points to coincide with the corresponding positions,
\begin{equation}
\left\{ \begin{array}{c}
x(s_{i-1})=x_{i-1}\\
x(s_{i})=x_{i}\\
x(s_{i+1})=x_{i+1}\end{array}\right.,
\end{equation}
This represents a system of equations for $a_{k}$, and there is an
analogous system for $b_{k}$. The values of these coefficients give
the explicit form of the functions $x(s)$ and $y(s)$ in the vicinity
of each point $s_{i}$ that can be used to find its derivatives up to
the second order. We note that the force on a point is different
from the force on an element $dl$ by a factor $u(s)$.


\section{\label{sec:Results}Validation}

The method described in this paper is intended as methodological;
there are no experimental results available for a two dimensional
system. The results we give below, demonstrate that the method
behaves in a manner which is consistent with the expected behaviour
of a two dimensional cell.  The possible extension of the method to
three dimensions is discussed in section~\ref{sec:Conclusion}.

The simulations in this section were run on D2Q9 $100\times100$
lattice with the periodic boundary conditions. The basis velocity
vectors of the D2Q9 lattice and corresponding values of $t_i$ are
presented in the table~\ref{tab:1}.  The primitive cell used for
averaging the velocities, and distributing the forces, was taken to
be a primitive square cell of the lattice whose corners are lattice
nodes. The value of the relaxation time $\tau=0.8$ was used in the
simulations. The surface tension coefficient and the preferred
curvature of the membrane were set to zero. The equilibrium distance
between the membrane points $\Delta s=3.2$ was used, and a typical
membrane included $50-100$ points.

\begin{table}[ht]
\caption{
\label{tab:1}
The basis velocity vectors of the D2Q9 lattice and corresponding values of $t_i$.
}
\begin{indented}
\item[]\begin{tabular}{ccc}
\br
$i$&$\mathbf{c}_i$&$t_i$\tabularnewline
\mr
0 &  (0,0)  & 4/9  \\
1 &  (1,0)  & 1/9  \\
2 &  (0,1)  & 1/9  \\
3 & (-1,0)  & 1/9  \\
4 &  (0,-1) & 1/9  \\
5 &  (1,1)  & 1/36 \\
6 & (-1,1)  & 1/36 \\
7 & (-1,-1) & 1/36 \\
8 &  (1,-1) & 1/36 \\
\br
\end{tabular}
\end{indented}
\end{table}

\begin{figure}
\begin{center}
\epsfbox{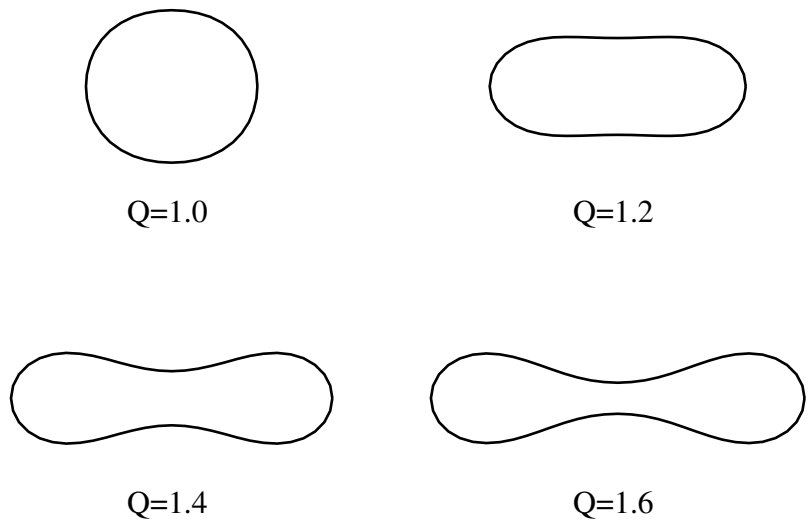}
\end{center}
\caption{
\label{fig:2}
Equilibrium shapes for different values of the parameter
$Q$ defined by (\ref{eq:Q}).
}
\end{figure}

Figure~\ref{fig:2} shows the equilibrium shapes of the vesicle for
different values of the parameter $Q$ defined as
\begin{equation}
Q=L_{0}/L_{c},
\label{eq:Q}
\end{equation}
$L_{c}$ being the length of the circular membrane with the same
enclosed area. The parameters are: $\kappa=0.05$, $\alpha=0.01$,
$\beta=0.2$. To provide the slight initial asymmetry, the initial
shape of the droplet was an ellipse with half-axes $20\pm0.2$.

\begin{figure}
\begin{center}
\epsfbox{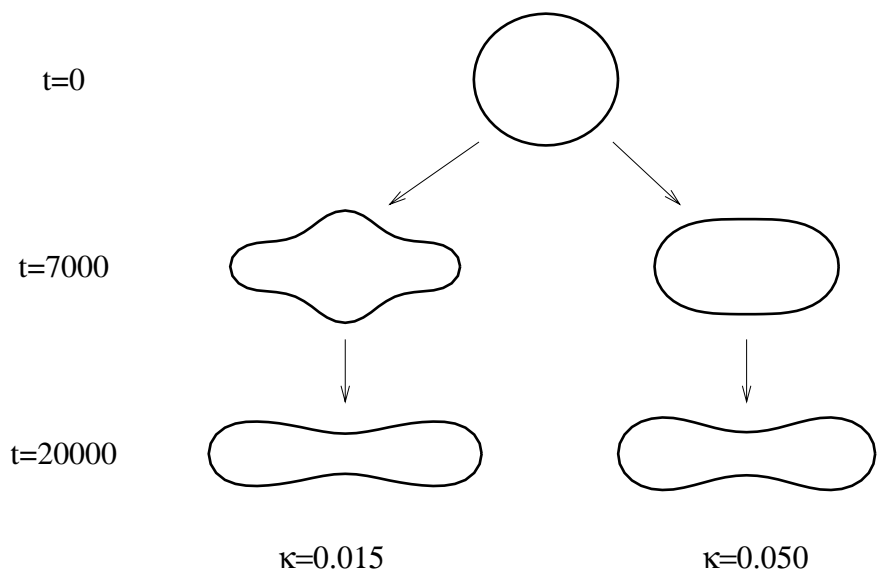}
\end{center}
\caption{
\label{fig:3}
Dynamics for different values of the bending rigidity
$\kappa$.
}
\end{figure}

\begin{figure}
\begin{center}
\epsfbox{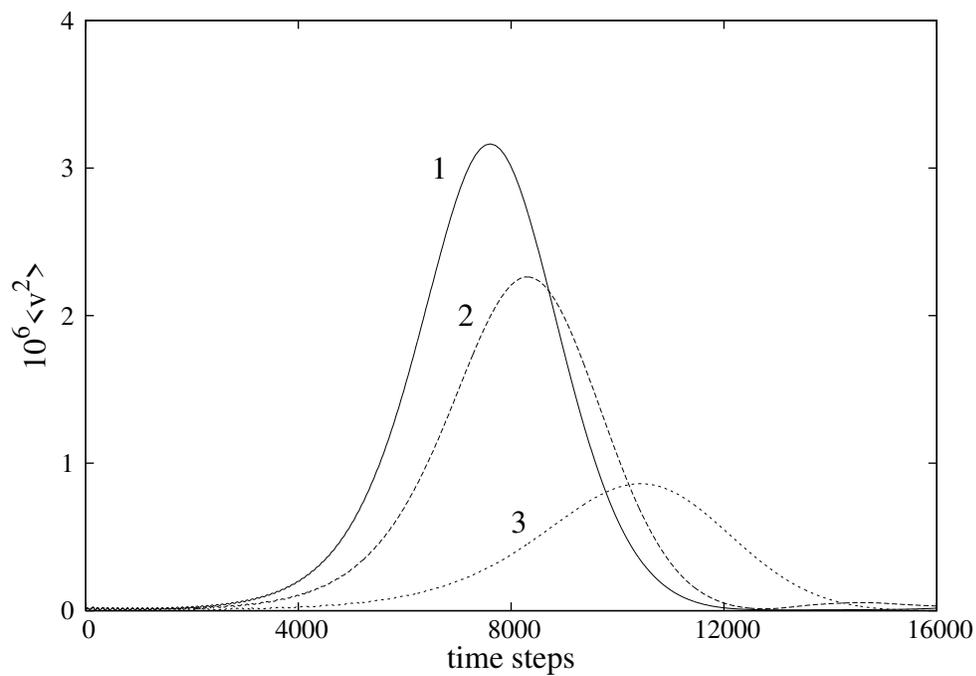}
\end{center}
\caption{ \label{fig:4} Time dependence of the mean square LB
velocity for $\kappa=0.025$ (curve 1), $\kappa=0.05$ (curve 2),
$\kappa=0.1$ (curve 3).
}
\end{figure}

It should be noted that the same equilibrium shapes can be obtained
without the including the effect of the embedding fluid; this was
confirmed as one of the validations of the simulation. However, the
inclusion of the LB fluid is necessary in order to recover the
dynamics of the relaxation to the equilibrium shape, as is shown in
Figure~\ref{fig:3} for $Q=1.4$ and different values of the bending
rigidity $\kappa$. The value of $\kappa$ also influences the rate of
relaxation which is demonstrated in Figure~\ref{fig:4} by the time
dependence of the mean square velocity of the LB fluid for different
values of $\kappa$.

\begin{figure}
\begin{center}
\epsfbox{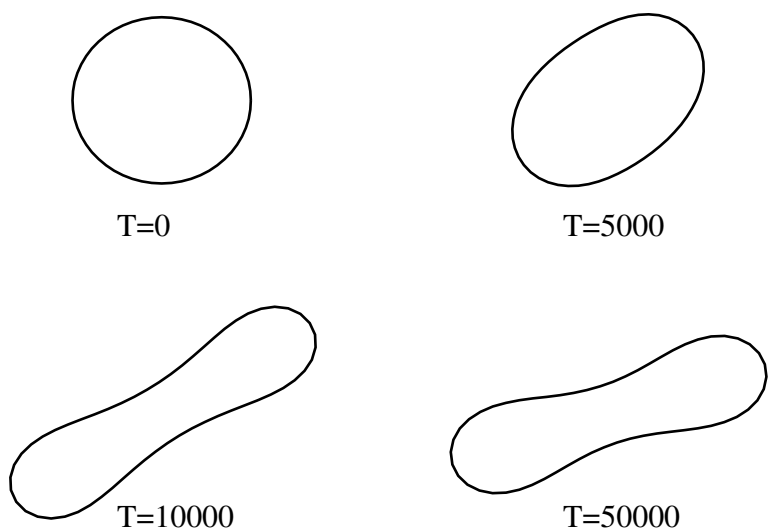}
\end{center}
\caption{
\label{fig:6}
Dynamics of the membrane in shear flow.
}
\end{figure}

\begin{figure}
\begin{center}
\epsfbox{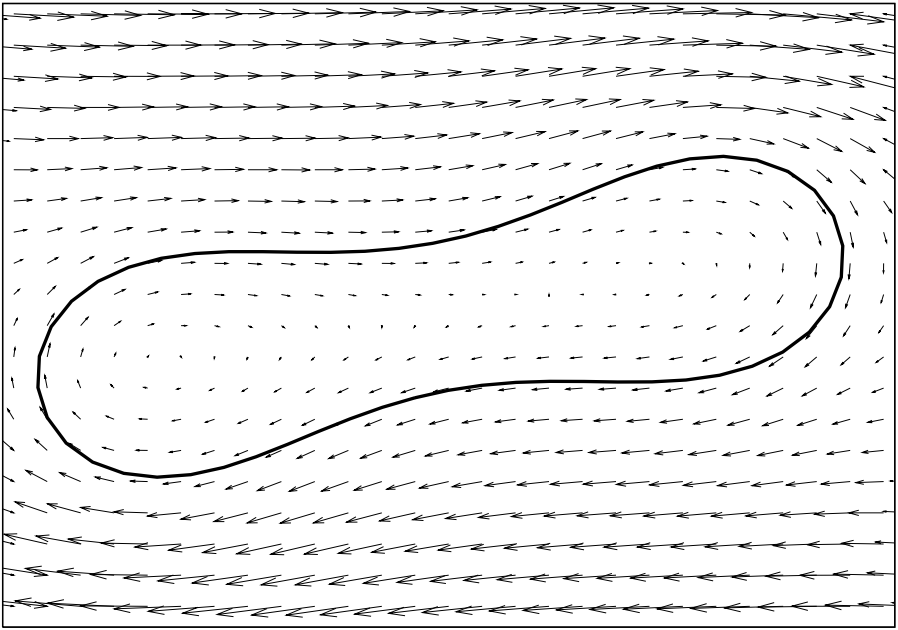}
\end{center}
\caption{
\label{fig:5}
The velocity field of the LB fluid under shear.
}
\end{figure}

To investigate the behaviour of the membrane in the flow, a
simulation was undertaken in which shear flow was applied to an
initially spherical membrane. Figure~\ref{fig:6} depicts the time
evolution of the shape of the droplet, and Figure~\ref{fig:5} shows
the velocity field of the LB fluid. Apart from the imposed shear,
the parameters of the simulation are the same as in
Figure~\ref{fig:2} with the parameter $Q$ was taken to be equal to
$1.4$.  In the final steady state the membrane rotates with the
fluid; however, this behaviour is expected only to occur in a
two-dimensional system.


\section{\label{sec:Conclusion}Conclusion}

A method has been described for embedding a deformable membrane into
a LB fluid and results presented which validate the approach in two
dimensions. The method does not take into account thermal
fluctuations. If the fluctuations are small they simply result in a
renormalisation of the bending rigidity coefficient
(Palmer and Morse 1996), and no further modification of the method
is necessary. Large fluctuation can be taken into account by adding
the random stress to the LB algorithm (Cates \etal 2004).

The method could be generalised to three dimensions by employing the
expression for the force due to bending rigidity of the
two-dimensional membrane based on a corrected version  of the result
derived by (Stelitano and Rothman 2000) (see comment after
equation (\ref{eq:FV})), and the method for calculating the
curvature of the triangulated surface which has been developed by
(Hamman 1993). Note that a grid would need to be
created for the equilibrium shape of the membrane prior to LB
simulation. This could be achieved by the direct numerical
minimisation of the free energy of the membrane. After the
minimisation, the equilibrium distances between points would be
changed and the details of the area contributions would therefore
need to be modified appropriately; work is currently in progress to
implement such a three dimensional scheme.

One possible application of the technique described in this paper
is a more accurate representation of the flow of blood cells in confined
geometries and in which membrane elasticity effects are taken into
account.


\References

\item[] Ahlrichs P and D\"unweg B 1998 {\it Int. J. Mod. Phys. C}  {\bf 9} 1429--38

\item[] Alberts B \etal 2002 {\it Molecular Biology of the Cell} (New York: Garland Science) p~1549

\item[] Canham P B 1970 {\it J. Theor. Biol.} {\bf 26} 61--81

\item[] Cates M E \etal 2004 \JPCM {\bf 16} S3903--15

\item[] Evans E 1974 {\it Biophys. J.} {\bf 14} 923--31

\item[] Hamman B 1993 {\it Computing Suppl.} {\bf 8} 139--53

\item[] Helfrich W 1973 {\it Z. Naturforsch.} {\bf 28c} 693--703

\item[] Lipowsky R 1991 {\it Nature} {\bf 349} 475--81

\item[] Lishchuk S V and Care C M 2005 {\it Phys. Rev. E} submitted

\item[] Palmer K M and Morse D C 1996 \JCP {\bf 105} 11147--74

\item[] Stelitano D and Rothman D H 2000 {\it Phys. Rev. E} {\bf 62} 667--80

\item[] Succi S 2001 {\it The Lattice Boltzmann Equation for Fluid Dynamics and Beyond}
(New York: Oxford University Press) p~288

\endrefs


\newpage

\end{document}